# Alternatives to the ROC Curve AUC and C-statistic for Risk Prediction Models

Ralph H. Stern
Division of Cardiovascular Medicine
Department of Internal Medicine
University of Michigan
Ann Arbor, Michigan

stern@umich.edu

## ABSTRACT

Assessment of risk prediction models has primarily utilized measures of discrimination, the ROC curve AUC and C-statistic. These derive from the risk distributions of patients and nonpatients, which in turn are derived from a population risk distribution. As greater dispersion of the population risk distribution produces greater separation of patient and nonpatient risks (discrimination), its parameters can be used as alternatives to the ROC curve AUC and C-statistic. Here continuous probability distributions are employed to develop insight into the relationship between their parameters and the ROC curve AUC and C-statistic derived from them.

The ROC curve AUC and C-statistic are shown to have a straight-line relationship with the SD for uniform, half-sine, and symmetric triangular probability distributions, with slight differences in the slope: AUC $\approx$ 1/2+0.28 SD/(mean(1-mean)). This also characterizes the beta distribution over the same range of SD's. But at larger beta distribution SD's the plot of AUC versus SD deviates downward from this straight-line relationship, approaching the ROC curve AUC and SD of a perfect model (AUC=1, SD= $\sqrt{mean(1-mean)}$).

Similar simple relationships can be derived for the overlap measure, Youden index, and Gini coefficient.

Unlike the ROC curve AUC and C-statistic, calculation of the mean squared error or Brier score is the same for any distribution: mean(1-mean)-SD$^2$, consistent with published decompositions of perfectly calibrated models.

A simpler and more intuitive discrimination metric is the coefficient of discrimination, the difference between the mean risk in patients and nonpatients. This is SD$^2$/(mean(1-mean)), which is also the same for any distribution.

Since estimating parameters or metrics discards information, the population risk distribution should always be presented. As the ROC curve AUC and C-statistic are functions of this distribution's parameters, the parameters represent simpler, intuitive alternatives to these discrimination metrics. Among discrimination metrics, the coefficient of discrimination provides a simple, intuitive alternative to the ROC curve AUC and C-statistic.

The assessment of risk prediction models has largely been focused on measures of discrimination, as these had previously been used in diagnosis. But prognosis differs from diagnosis in a fundamental way in that the risk distributions of patients and nonpatients are not independent; they are both fully determined by the population risk distribution. More disperse population risk distributions produce greater separation between the risk distributions of patients and nonpatients, i.e., discrimination. Thus, risk prediction models can be assessed graphically by presentation of the population risk distribution as an alternative to the risk distributions of patients and nonpatients and/or the ROC curve derived from them. And they can be assessed mathematically by presentation of parameters of the population risk distribution as an alternative to measures of discrimination such as the ROC curve AUC or C-statistic.

Since the population risk distribution is the root of all graphical and numerical assessments, this paper explores the quantitative relationship between its parameters and derived discrimination measures.

## THE POPULATION RISK DISTRIBUTION

To develop insight into the relation between parameters and measures of discrimination, parametric continuous probability distributions have been utilized. These perfectly calibrated models are not affected by the noise of real-world data. And in many instances analytical solutions are available that permit discrimination and other metrics to be expressed analytically as a function of the parameters of the population risk distribution.

## THE ROC CURVE AUC

The ROC curve presents the true positive versus false positive rate. Since both rates are calculated at the same threshold or cutoff, the ROC curve is a parametric curve. The area under this parametric curve is [1]:

$$AUC = \int_0^1 \frac{(1-r)f(r)}{(1-mean)} \left( \int_r^1 \frac{xf(x)}{mean}\, dx \right) dr$$

r is absolute risk,
f(r) is the probability density function of r for the population,
mean is the population risk,
(1-r) f(r)/(1-mean) is the probability density function of r for nonpatients,
r f(r)/mean is the probability density function of r for patients (x above is a dummy variable)

As discussed previously, this integral expression provides a quantitative definition for the qualitative concept of discrimination and a characterization of the C-statistic as a Monte Carlo integration of this expression. [1]

The simplest continuous probability distribution is the uniform distribution. Although such a distribution may seem unrealistic, it will be shown to be informative. Using the conventional parameterization with a as the lower bound and b as the upper bound:

$$AUC = \int_a^b \frac{1-r}{(b-a)(1-mean)} (\int_r^b \frac{x}{(b-a)mean} dx) dr$$

However, for our purposes, we will reparametrize in terms of the mean ((a+b)/2) and delta (b-a):

$$AUC = \int_{mean-\frac{delta}{2}}^{mean+\frac{delta}{2}} \frac{1-r}{delta(1-mean)} (\int_r^{mean+\frac{delta}{2}} \frac{x}{delta\ mean} dx) dr$$

which can be integrated to give:

$$AUC = \frac{1}{2} + \frac{delta}{12\ mean(1-mean)}$$

As the standard deviation (SD) for a uniform distribution is delta/ (2 $\sqrt{3}$),

$$AUC = \frac{1}{2} + \frac{\sqrt{3}}{6} \frac{SD}{mean(1-mean)}$$

For two other parametric distributions, the coefficients of $\frac{SD}{mean(1-mean)}$ are:

$\frac{\pi}{8\sqrt{\pi^2-8}}$ for a half-sine distribution

$\frac{7\sqrt{6}}{60}$ for a symmetric triangular distribution

The decimal values of the coefficients for these three parametric distributions are similar at 0.289, 0.287, and 0.286, so a general AUC approximation is:

$$AUC \approx \frac{1}{2} + 0.28 \frac{SD}{mean(1-mean)}$$

A more realistic distribution is the beta distribution. Although a simple analytic expression for the AUC cannot be derived, numerically exploring the relationship between AUC, SD, and mean over the range of SD's possible with uniform distributions with the same means shows similar relationships with the coefficients ranging from 0.271 for a beta distribution with a mean of 0.01 to 0.289 with a mean of 0.5. Since these coefficients are nearly the same as for the three simple parametric distributions, graphs of the AUC vs SD for the beta distribution and uniform distribution are nearly superimposable over the shared range of SD's. (The SD of the uniform distribution is limited by its limited range.) Thus, there is a straight-line relationship between SD and AUC for selected simple parametric distributions, and even for the beta distribution over the shared range of SD's.

The Figure shows the ROC curve AUC as a function of SD for both beta and uniform distributions with means of 0.01, 0.05, 0.1, 0.2, and 0.5 over a broad range of selected SD's. They are about the same over the shared range of SD's, up to SD's corresponding to ROC curve AUC's of about 0.75.

The maximal SD for a risk distribution is that of a perfect model, which would produce two risk categories, one with an absolute risk of 0 (consisting entirely of nonpatients) with a density equal to 1-mean and one with an absolute risk of 1 (consisting entirely of patients) with a density equal to the mean. The SD of a perfect model is:

$$\sqrt{(1-mean)(0-mean)^2 + mean(1-mean)^2} = \sqrt{mean(1-mean)}$$

For risk distributions with means of 0.01, 0.05, 0.1, 0.2, and 0.5, the maximal SD's are 0.0995, 0.2179, 0.3, 0.4, and 0.5, respectively. As is well known, the ROC curve AUC for a perfect model would be 1. As the SDs of the beta distribution increase beyond the range of SD's shared with the uniform distribution, the AUC versus SD relationship progressively deviates downward from the straight-line relationship to approach the AUC of 1 and SD of $\sqrt{mean(1-mean)}$ of a perfect model.

## THE COEFFICIENT OF DISCRIMINATION

A simpler and more intuitive measure of discrimination is the coefficient of discrimination, which is the difference in absolute risk between patients and nonpatients. [2]

Begg et al. have shown that the standardized incidence ratio (risk of second cancer in a patient/risk of first cancer in the population) is $1+CV^2$, where CV is the coefficient of variation of the risk distribution. [3] Adopting the above nomenclature:

$$\frac{mean\ risk\ in\ patients}{mean} = 1 + \left(\frac{SD}{mean}\right)^2$$

which can be derived from:

$$\frac{mean\ risk\ in\ patients}{mean} = \frac{\int r\frac{rf(r)}{mean}}{mean} = \frac{\int r^2\ f(r)\ dr}{mean^2} = \frac{E[r^2]}{mean^2} = \frac{mean^2+Var[r]}{mean^2} = 1 + \left(\frac{SD}{mean}\right)^2$$

So:

$$mean\ risk\ in\ patients = mean + \frac{SD^2}{mean}$$

Similarly:

$$mean\ risk\ in\ nonpatients = mean - \frac{SD^2}{1-mean}$$

Thus:

$$coefficient\ of\ discrimination = \left(mean + \frac{SD^2}{mean}\right) - \left(mean - \frac{SD^2}{1-mean}\right)$$

$$= \frac{SD^2}{mean(1-mean)}$$

As with the ROC curve AUC, the coefficient of discrimination is a simple function of SD and mean of the population risk distribution, but unlike the ROC curve AUC, this function is the same for any probability distribution.

## THE BRIER SCORE

The Brier score is the mean squared error. With real world data, it is a measure of both discrimination and calibration, but with parametric distributions it is only a measure of discrimination.

$$Brier\ score = \int r\, f(r)(1-r)^2 + (1-r)f(r)(0-r)^2 dr$$

$$= \int (r - 2r^2 + r^3 + r^2 - r^3) f(r) dr = \int (r - r^2) f(r) dr = \text{mean-E}(r^2)$$

$$= mean - (mean^2 + Var(r)) = mean(1 - mean) - SD^2$$

Perfect models have a coefficient of discrimination of 1 and a Brier score of 0. As with the ROC curve AUC, the Brier score is a simple function of SD and mean of the population risk distribution, but unlike the ROC curve AUC, this function is the same for any probability distribution.

## OTHER METRICS

There is an abundance of other metrics in the literature that can also be shown to be simple functions of the population risk distribution parameters. For example:

Overlap measure: $1 - coefficient \frac{SD}{mean(1-mean)}$

Coefficients:

- Uniform $\frac{\sqrt{3}}{4} = 0.433$
- Half-sine $\frac{\pi-2}{2\sqrt{\pi^2-8}}$ = 0.417
- Triangular $\frac{1}{\sqrt{6}}$ = 0.408

Youden index: $1 - Overlap\ measure$

Gini coefficient: $coefficient\ \frac{SD}{mean}$

Coefficients:

- Uniform $\frac{1}{\sqrt{3}}$ = 0.577
- Half-Sine $\frac{\pi}{4\sqrt{\pi^2-8}} = 0.574$

Triangular $\frac{7\sqrt{6}}{30} = 0.572$

DISCUSSION

Because of the focus on discrimination, if there is a graphical presentation in publications, most commonly it is the ROC curve and/or the risk distributions of patients and nonpatients. Only the latter is mentioned in the TRIPOD guidance. [4]

The population risk distribution can be presented as a risk distribution curve (a probability density function), a cumulative risk distribution curve, or a predictiveness curve [5] (a plot of risk versus the cumulative distribution function). Pepe wrote "Displaying risk distributions is a fundamental step in evaluating the performance of a risk prediction model, a step that is often overlooked in practice." [6]

Even though these displays allow one to appreciate the location, the spread, and the shape of the population risk distribution, these distributions rarely appear in publications. Kent et al. have published several examples.[7] As estimating parameters or metrics discard the wealth of information included in the population risk distribution, presentation of these estimates alone should be discouraged. Inspection of the distribution allows an immediate impression of how well the model performs at separating the population into risk subgroups differing as much as possible from the population risk, which is the basis for their clinical benefit, a more efficient allocation of preventive measures dependent on risk level. And when comparing newer and older models, it allows an immediate impression of the difference. Deciding whether use of a model or improvement in a model will have clinical benefits requires consideration of factors external to the model. In many instances, these visual evaluations may provide a sufficient basis for deciding on the need for additional analyses.

When metrics are presented, they are usually discrimination metrics (ROC curve AUC and C-statistic) that describe the relationship between the risk distributions of patients and nonpatients. For risk prediction models, since the population risk distribution fully determines the risk distributions of patients and nonpatients, parameters of the root population risk distribution are more fundamental than the usual metrics, in addition to being simpler and more intuitive.

Although the quantitative relationship between parameters and derived metrics were explored for a limited number of continuous probability distributions, the computations are primarily to provide insights, which are generally applicable. However, the beta distribution should be a reasonable approximation to many or most real-world population risk distributions.

Since estimating parameters or metrics discards information, the population risk distribution should always be presented. As the ROC curve AUC and C-statistic are functions of this distribution's parameters, the parameters represent simpler, intuitive alternatives to these discrimination metrics. Among discrimination metrics, the coefficient of discrimination provides a simple, intuitive alternative to the ROC curve AUC and C-statistic.

## FIGURE LEGEND

ROC curve AUC as a function of SD for both the beta distribution (points) and uniform distribution (lines) with means of 0.01, 0.05, 0.1, 0.2, and 0.5.

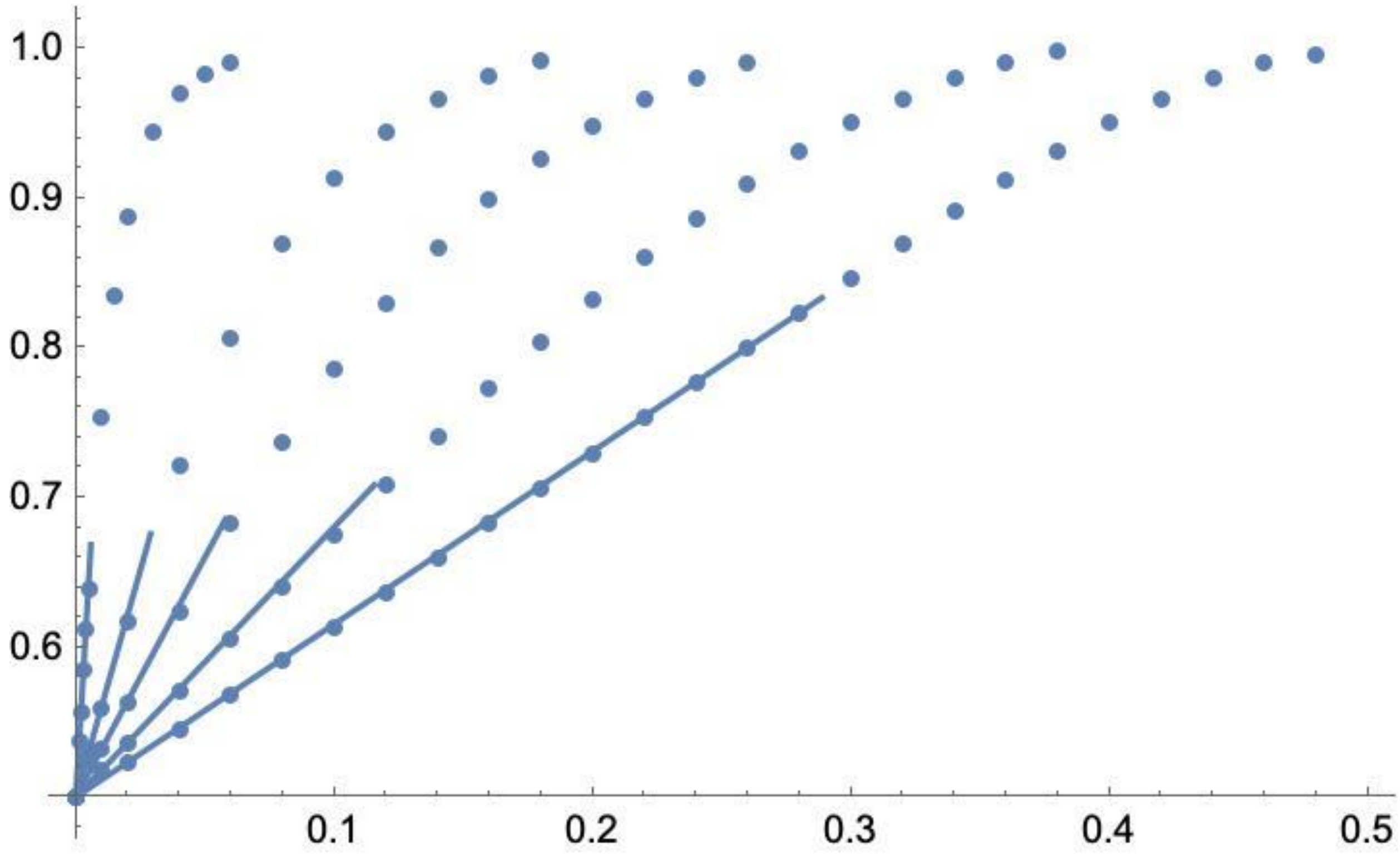